\documentclass[10pt, conference]{IEEEtran}
\IEEEoverridecommandlockouts

\usepackage{cite}
\usepackage{amsmath,amssymb,amsfonts,amsthm}
\usepackage{algorithmic}
\usepackage{graphicx}
\graphicspath{{fig}{./fig/}}
\usepackage{textcomp}
\usepackage{xcolor}
\usepackage{hyperref} %
\usepackage{comment}
\usepackage{footmisc}
\usepackage{subfigure}
\usepackage{flushend} %
\usepackage[linesnumbered, vlined, ruled]{algorithm2e}
\usepackage{soul} %

\usepackage{etoolbox}
\AtBeginEnvironment{equation}{\fontsize{9}{12}\selectfont}
\AtBeginEnvironment{align}{\fontsize{9}{12}\selectfont}

\SetCommentSty{mycommfont}

\hypersetup{
    colorlinks=true,  %
    linkcolor=black,  %
    citecolor=black,  %
    urlcolor=black    %
}

\def\BibTeX{{\rm B\kern-.05em{\sc i\kern-.025em b}\kern-.08em
    T\kern-.1667em\lower.7ex\hbox{E}\kern-.125emX}}

\begin{document}

\title{Real-Time AI Acceleration on a 2D Spatial Dataflow Architecture}

\author{\IEEEauthorblockN{Shriram Raja, Anton Njavro, Richard West}
    \IEEEauthorblockA{Department of Computer Science
                      \\Boston University
                      \\ Boston, MA}
    \{shriramr, njavro, richwest\}@bu.edu
}

\maketitle

\begin{abstract}

Machine Learning (ML) is increasingly being integrated into real-time systems,
which makes the timing analysis of ML accelerators necessary. However, the most
popular commercial accelerator, the Graphics Processing Unit (GPU), remains
difficult to analyze as several aspects of its behavior, including scheduling of
threads involved in single-instruction multiple-thread computations and memory access latency, are not fully
documented.

Spatial dataflow accelerators present an interesting alternative to GPUs, as they inherently support compute and memory partitioning. Data movement
and computation locality are explicitly controlled by the programmer. In this
work, we consider the Tenstorrent Blackhole, which is based on a commodity RISC-V architecture. It
consists of a 2D grid of cores connected by Networks-on-Chip (NoCs), supporting
a fully open software runtime. We profile the Blackhole platform and develop a
worst-case execution time task model as a Homogeneous Synchronous Dataflow
Graph. We extend this model to a preliminary analysis of multi-core
workloads by considering NoC-related delays. Finally, we evaluate our model across
$12{,}567$ task configurations and workloads. Results show that the estimate
derived from our model lies above the observed worst case in every
configuration. For single-core workloads, the overhead is less than 100\% in
most cases, with a maximum of less than 200\%, and for multi-core workloads, it is less than 500\% using a pessimistic NoC contention analysis. This paper lays the foundations for
real-time computing on a promising open accelerator design.
\end{abstract}


\section{Introduction}\label{sec:introduction}
Machine Learning (ML) inference is increasingly deployed in safety-critical
systems such as self-driving cars, necessitating the analysis of AI accelerators
that are used in such systems. A substantial body of work addresses one half of
that problem: real-time scheduling on accelerator-based
platforms~\cite{ai_accelerator_rt_survey}. Scheduling analysis, however, takes a
per-task worst-case execution time (WCET) as its input, and that input is
difficult to obtain on the most popular commercial AI accelerator, the Graphics
Processing Unit (GPU). Several key aspects of its behavior, including how work
is dispatched to the compute units, and how operands reach them through the
memory hierarchy, are largely undocumented, so WCET analysis has to consider the
reverse-engineering of those
properties~\cite{gpu_sched_hidden_details,dissect_cuda,demystifying_gpu} or make
assumptions about them~\cite{wcet_gpu_warp}. Even when the scheduling aspects of
GPUs are made predictable through compute
partitioning~\cite{libsmctrl,nvtaskset,nvidia_mig,nvidia_green_ctxt}, little
work has been done to address the predictable partitioning of the memory
hierarchy. Accounting for the shared memory hierarchy is difficult as the
arbitration rules remain undocumented. Reverse-engineering is itself made
difficult on such platforms as the software stack also remains closed source. 

The Tenstorrent Blackhole presents an interesting alternative. Since its
software stack is fully open, determining the behavior of the platform by
profiling becomes tractable. In this work, we choose it as a representative
spatial dataflow architecture. Typically organized as a two-dimensional grid of
cores connected by Networks-on-Chip (NoCs), such architectures alleviate the
pressure on a global shared memory~\cite{TileLoom}, and offer the potential for
spatio-temporal partitioning of tasks. The Blackhole is a commodity RISC-V
accelerator with a 2D grid of ``Tensix'' and other specialized cores
connected by two NoCs. There is no shared cache hierarchy for data, no hardware
prefetcher, and no runtime-managed migration between memories. Thus, all data
movement is explicitly managed by kernels running on each Tensix core. This
simplifies worst-case timing analysis as data movement is determined from the
kernel source code rather than inferred from a cache model. The programming
model separates read, compute and write stages, similar to the
acquisition-execution-restitution three-phase approach~\cite{PREM,AER}.

In prior work focusing on the Tenstorrent Blackhole, researchers have performed
preliminary benchmarking~\cite{blackhole_bm}, built compilers that plan dataflow
and schedules across its core grid~\cite{TileLoom,loom}, allocated cores among
concurrent jobs~\cite{SwiftNPU}, and ported GPU code~\cite{cuda_migrate_to_TT,HetGPU}. However, none of them analyze the 
worst-case execution time of the architecture. On an older hardware generation, Gandhi et al. predict the typical rather than worst-case 
execution time of convolution~\cite{tt_wormhole_pred_exec}, but do not consider
any other workloads. In this work, we evaluate the suitability of the Tenstorrent Blackhole for
real-time applications, by deriving a model of worst-case
execution time for a task, defined as all the code running on one or more Tensix
cores.

{\bf Contributions:}
\begin{itemize}
  \item We profile the Tenstorrent Blackhole architecture and describe the
  properties of the platform in sufficient detail to facilitate timing
  analysis.

  \item We use these properties to model a task as a Homogeneous Synchronous
  Dataflow Graph and derive a WCET bound for a single-core task in the absence
  of NoC contention, given upper bounds on the model parameters. 

  \item We then extend the WCET bound to multi-core workloads by accounting for
  the effect of inter-core NoC contention using the closest existing
  analysis~\cite{BATA}.
  
  \item We finally evaluate the model using observed maxima, which gives WCET
  estimates~\cite{wcet_problem}, and show that, in all configurations, our
  estimate is greater than the observed worst case. For single-core workloads,
  the overhead is less than 100\% in most cases, with a maximum below 200\%, and
  for workloads up to 32 cores, the maximum overhead is less than 500\%. Relying
  on observed maxima is not a fundamental limitation of our
  approach, as the bound holds for any proven upper bounds on its parameters,
  such as those from static analysis.
\end{itemize}

The rest of the paper is organized as follows. Section~\ref{sec:bg} provides
background on the Blackhole architecture and its programming model.
Section~\ref{sec:tt_prof} describes the properties of the platform determined by
profiling. Section~\ref{sec:timing_model} introduces the task model, derives the
worst-case execution time of single-core tasks, and extends it to multi-core by
accounting for NoC contention. Section~\ref{sec:exp_eval} evaluates our model.
Section~\ref{sec:rel_work} discusses related work, and
Section~\ref{sec:conclusion} concludes the paper.

\section{Background}\label{sec:bg}
\begin{figure}[htbp]
  \centering
  \includegraphics[width=0.75\linewidth]{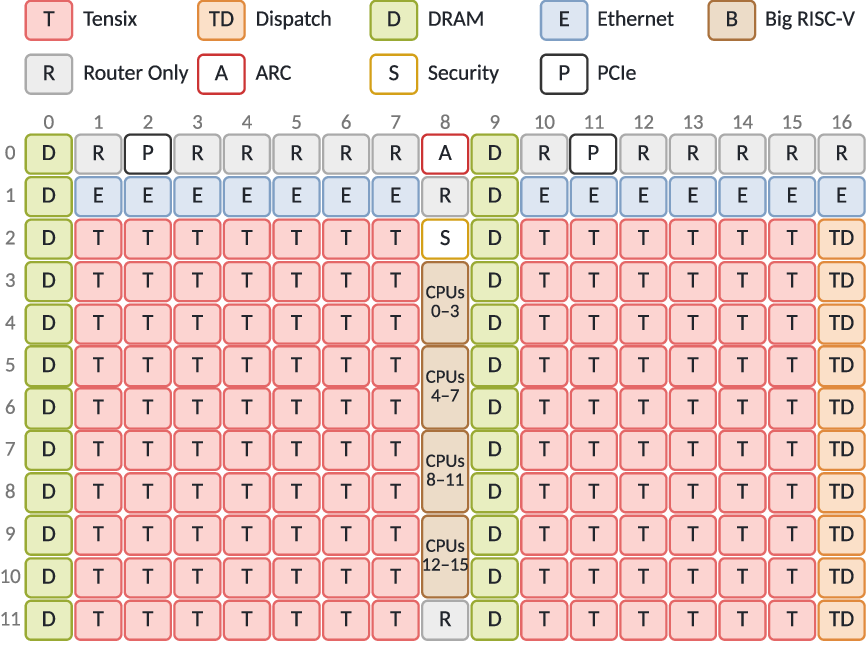}
  \caption{The Blackhole p150a architecture.}
  \label{fig:coremap}
\end{figure}

Tenstorrent accelerators are dataflow machines built as a two-dimensional grid
of heterogeneous cores connected by two Networks-on-Chip (NoCs). We use a
Blackhole p150a board, whose grid (shown in Figure~\ref{fig:coremap}) is $17$
columns by $12$ rows and holds $140$ Tensix cores (compute nodes), $24$ DRAM
cores that expose a total of $32$~GiB of GDDR6 memory, $14$ Ethernet cores for
board-to-board links, two PCI Express cores, four clusters of general-purpose
(``big'') RISC-V cores~\cite{tt_isa_docs} which are capable of running
Linux~\cite{tt_linux}, and specialized cores for security and firmware (ARC).
Note that other spots on the grid only hold routers for the NoCs.

A program reaches the device either by the host writing binaries and arguments
straight into device memory and polling for completion ({\it slow dispatch}), or
by enqueuing commands that dispatcher cores on the device consume ({\it fast
dispatch})~\cite{tt-metal}. We use fast dispatch throughout, since serializing
the host against the device defeats the purpose of an accelerator. 
This reserves ten cores (column 16) for processing of the command queue, as
shown in Figure~\ref{fig:coremap}.

The on-chip clock is generated by a phase-locked loop programmed by firmware
running on a management core (ARC), which also runs a dynamic voltage and
frequency scaling arbiter that picks a frequency between a minimum and a
maximum. The host only sends a busy or idle bit, for which the arbiter targets
$1350$~MHz and $800$~MHz, respectively. All Tensix cores read the shared
frequency from the ARC core and maintain per-core cycle counters. These per-core
counters are read to obtain all the timing measurements in this work.

\subsection{Tensix Core}

\begin{figure}[htbp]
  \centering
  \subfigure[]{%
    \includegraphics[width=0.42\linewidth]{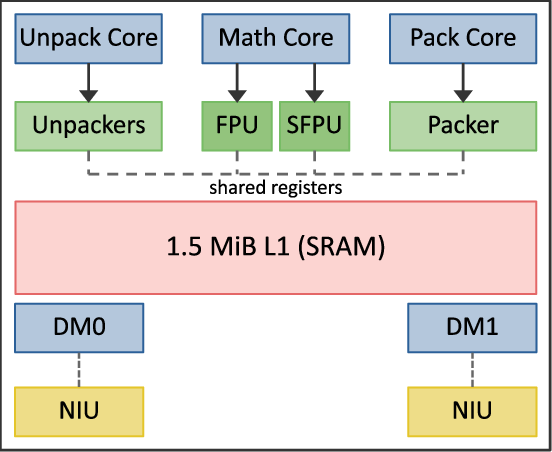}%
    \label{fig:tensix_core}}
  \hfill
  \subfigure[]{%
    \includegraphics[width=0.42\linewidth]{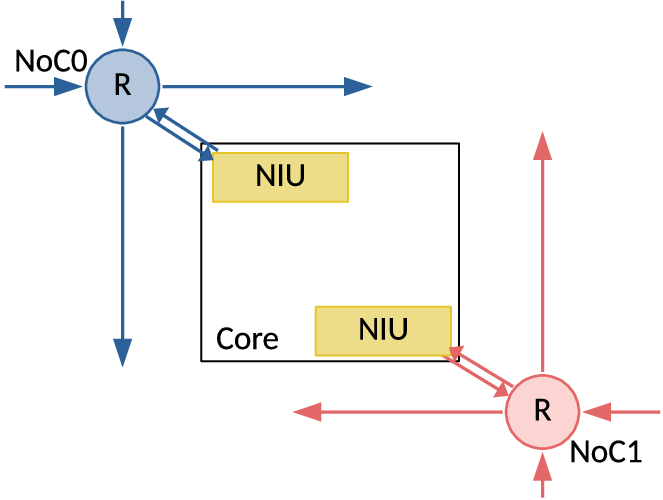}%
    \label{fig:noc_niu}}
  \caption{(a) Tensix core: five baby RISC-V cores (in blue) and the
  coprocessors they drive. (b) NoC directions, the NIU and the router on one
  core.}
\end{figure}

A Tensix core, shown in Figure~\ref{fig:tensix_core}, contains five ``baby''
RISC-V cores, their associated coprocessors, and $1.5$~MiB of SRAM called the L1.
The baby RISC-V cores are small, in-order, single-issue $32$-bit cores
implementing RV32IM with a handful of extensions~\cite{tt_isa_docs}. Each has a
few kilobytes of private local RAM for its stack and frequently accessed
variables. The baby RISC-V cores do not perform any matrix computation or bulk
data movement by themselves. Instead, they issue instructions to a coprocessor
and continue. Thus their instruction stream is a mixture of ordinary RISC-V
instructions and Tensix instructions that are forwarded to the coprocessor. The
cores have no context-switching capabilities~\cite{tt-metal} and run a single
function to completion.
 
Two baby RISC-V cores are designated as Data-Movement (DM) cores, and are
capable of driving either one or both of the NoC Interface Units (NIU) to move
data in and out of the L1 memory of the Tensix core. 
The other three baby RISC-V cores together perform the compute. 
The Unpack core controls the unpackers, which move input data from the L1 into
registers and convert its format if required. The Math core controls two
processing units: a Floating Point Unit (FPU) for simple matrix operations
(e.g., multiplication, addition and transposing), and a Special Function
Processing Unit (SFPU) for other operations (e.g., trigonometric operations).
The Pack core drives the packer, which moves results back to the L1. 

\subsection{Memory Subsystem}

{\bf L1:} The per-Tensix L1 holds all the data and code used by the processing
units on the Tensix core. It is banked, and access is arbitrated through a set
of ports, any of which can reach any bank. 

{\bf DRAM:} The $32$~GiB of GDDR6 is organized as eight partitions of
approximately $4$~GiB, with a DRAM controller per partition.\footnote{While
these partitions are called channels in Tenstorrent's documentation, they are
distinct from DRAM channels in the JEDEC sense.} Each controller
is reachable from three DRAM cores. All three DRAM cores expose an identical
view of their shared controller~\cite{tt_isa_docs}. 
Tensix cores are capable of accessing any DRAM partition through any of the
three associated DRAM cores on either NoC. This provides the programmer with a
high degree of control.

{\bf Caches:} The only caches on the device are the data and instruction caches
between each baby RISC-V core and the L1, and both are private to that RISC-V
core. The data cache is disabled by default~\cite{tt-metal}.
Neither the unpackers, the packer nor the NoCs pass through it, so no tensor
data is cached anywhere on the device. The instruction cache, by contrast, is
always in the path~\cite{tt_isa_docs}, but is accounted for by measuring the
cost of an operation with a cold cache. 

\subsection{Network-on-Chip}\label{subsec:noc}

The two NoCs, NoC0 and NoC1, are physically separate, connect all cores, and
flow in opposite directions. Each is a two-dimensional torus: the left and right
edges are joined, as are the top and bottom. 
A NoC0 packet travels right as necessary, turns at most once, and then travels
down, while a NoC1 packet travels up as necessary, turns at most once, and then
travels left. A NoC read involves a request to the destination and a
response, while a write involves sending data to the destination and waiting to
receive an acknowledgment. As each ring is unidirectional and requests cannot
switch between the NoCs, all reads and writes require a round-trip traversal of
the NoC from the requesting core, through the endpoint, and back.

On the Blackhole p150a, a full ring is $17$ hops along the horizontal ($x$) axis
and $12$ on the vertical ($y$) axis. For instance, a transaction with the source
at core $(x=1,y=2)$, and destination at $(14, 2)$ on NoC0 would involve the request
traveling $13$ hops, and the response traveling $4$ hops, making a total of
$17$ hops all along the $x$-axis. When the source and destination differ in both their $x$ and $y$
coordinates, e.g., cores $(2, 3)$ and $(11, 8)$, their round-trip distance is
the sum of the distances along both axes, i.e., $17 + 12 = 29$ hops. We refer to
the round-trip distance as $NoCrt$ in the rest of this paper. Even
though the one-way distance between two cores is different on each NoC as they
route packets in different directions, the $NoCrt$ is the same. 

Every packet originates and terminates at an NIU, and there
is one NIU per NoC on each core, hence two per Tensix core, as shown in
Figure~\ref{fig:noc_niu}. The data movement baby RISC-V core sends the required
number of requests to the NIU and issues a barrier to wait until the initiated
transactions finish. A packet is split into multiple ``flits'', each of which is
$64$~B large. A single NoC packet consists of one header flit and at most $256$
data flits, resulting in a maximum packet size of $16$~KiB. The number of flits
in a packet is an important factor to be considered when determining the
worst-case NoC transfer time in the presence of interference. There are sixteen
Virtual Channels (VCs) on each NoC, with support for static and dynamic VC
assignment.

\subsection{Programming Model}

{\bf Tile:} A Tensix program typically operates on $32 \times 32$-sized blocks
of data, called \emph{tiles}. Every compute operation consumes and produces
whole tiles.
As a tile's element count is fixed, its footprint is determined by its datatype
alone.
Data is either converted to a tile format (tilized) by the host before sending,
or is converted by the device code on arrival, at the cost of a compute
operation.

{\bf Software Stack:} Software for a Tenstorrent device is organized in
three levels~\cite{tt-metal}. At the top, models developed using ML frameworks are compiled for the target device. Below the model layer is the TT-NN library that provides operations, such as matrix multiplication and convolution, on which ML models are built. The lowest layer is the TT-Metalium runtime\footnote{Henceforth, we abbreviate ``TT-Metalium'' to simply ``Metalium''.}, which supports the construction of custom kernels that run on specific Tensix cores. We work
at the TT-Metalium level, since this is the level at which every quantity our model
needs
is named explicitly. 
While the software stack is updated frequently, the hardware is documented in
depth~\cite{tt_isa_docs} and the runtime is open source~\cite{tt-metal}, which
is what makes real-time analysis of this machine tractable.

{\bf Metalium Kernel:} The unit of code in a Metalium program is a {\it
kernel}: a source file, compiled by the host, that runs to completion on a
baby RISC-V core. A program for one Tensix core involves the creation of three
kernels: two data-movement kernels that run on the DM RISC-V cores and act as
reader and writer, and one compute kernel. 
The compute kernel is written as a single file, but compiled three times,
producing separate binaries for the Unpack, Math and Pack RISC-V cores. 
The kernels communicate with each other through circular buffers of predefined
depth (specified in tiles) and datatype. We refer to the buffer between the
reader and the compute kernels as the input buffer, and the buffer between the
compute and writer kernels as the output buffer. 
In each case, the producer waits for a free slot before pushing data into it,
either through a NoC transaction in the case of the reader kernel, or by writing
data from registers in the case of the compute kernel. The consumer waits for a
filled slot before popping the data out to a remote destination, through the NoC
in the case of the writer kernel, or into registers in the case of the compute
kernel. 

This programming model follows the
acquisition-execution-restitution approach~\cite{PREM,AER}.
There is no implicit data movement, as there is no shared cache hierarchy, no
hardware prefetcher and no runtime-managed migration between memories. The data
movement of a program is explicit in its code, and can be enumerated, attributed
to a specific source and destination, and accounted for, rather than inferred
from a cache model. The same holds for the compute stage, since a kernel specifies the
operations it performs and the number of tiles involved in them.

\section{Profiling the Tenstorrent Blackhole}\label{sec:tt_prof}
In this section, we detail the results of the profiling we performed on our
Blackhole to build a timing model. We used a Blackhole p150a with the release
$0.75.0$ of the Metalium stack with firmware version $19.8.1$.\footnote{Since
firmware version 19.5.0, two columns of the Tensix cores are marked inaccessible
to match devices shipped after January 2026 that do not have them. Hence the
number of worker Tensix cores on our setup is 110.}

\subsection{Remote Memory Access}\label{subsec:rma_bm}

To begin with, we benchmarked the remote memory access latency. We chose Tensix
core $(1, 2)$ as the source of the NoC transaction and accessed data on the L1
of all other (109) worker Tensix cores (non-dispatch) on the board. Each remote
core's L1 was accessed on both NoCs and the data size of each transaction was
varied from $16$~B (the alignment of the L1) to $1$~MiB, for a total of 10
sizes. One size of $1088$~B accounted for the tile size with BFP8 data types, which Tenstorrent refers to as {\tt Bfp8\_b}. The read and write latency was thus measured for a total of $109 \times 2
\times 10 = 2180$ configurations with each of them repeated for $15{,}000$
iterations. The experiment was repeated with a few other cores as reference and
it was noted that the observations made below remained valid. 

\begin{figure}[htb]
  \centering
  \includegraphics[width=\linewidth]{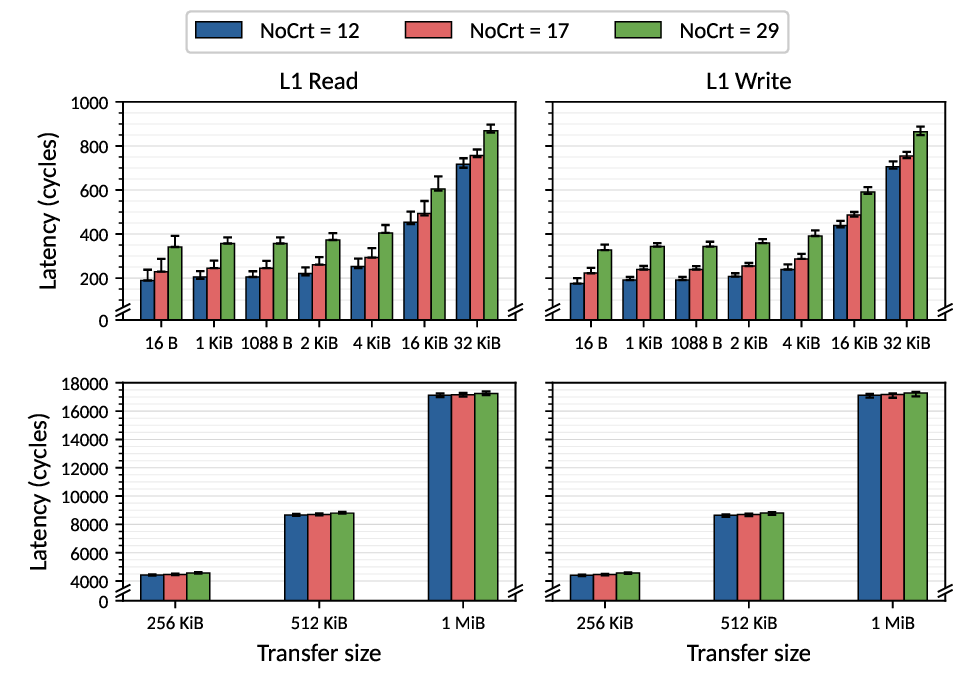}
  \caption{Remote L1 access latency at different sizes. Bars show the
  median and error bars the minimum and maximum.}
  \label{fig:l1_rw}
\end{figure}

{\bf Observation 1:} Remote memory access time depends on the size of data and
the round-trip distance ($NoCrt$) between the source and destination.

The read and write latencies obtained for all ten sizes are shown in
Figure~\ref{fig:l1_rw}. The measurements confirmed the assertion of the
documentation that the two NoCs are symmetrical. We also gained a key insight:
for a specific data size, the latency does not depend on the one-way distance,
but the round-trip distance between the reference core and the destination
Tensix core. As seen in the figure, latencies split into three different classes
based on $NoCrt$. This is potentially because both read and write transactions
involve round-trip traversal of a NoC (as stated in Section~\ref{subsec:noc}).

\begin{figure}[htb]
  \centering
  \includegraphics[width=\linewidth]{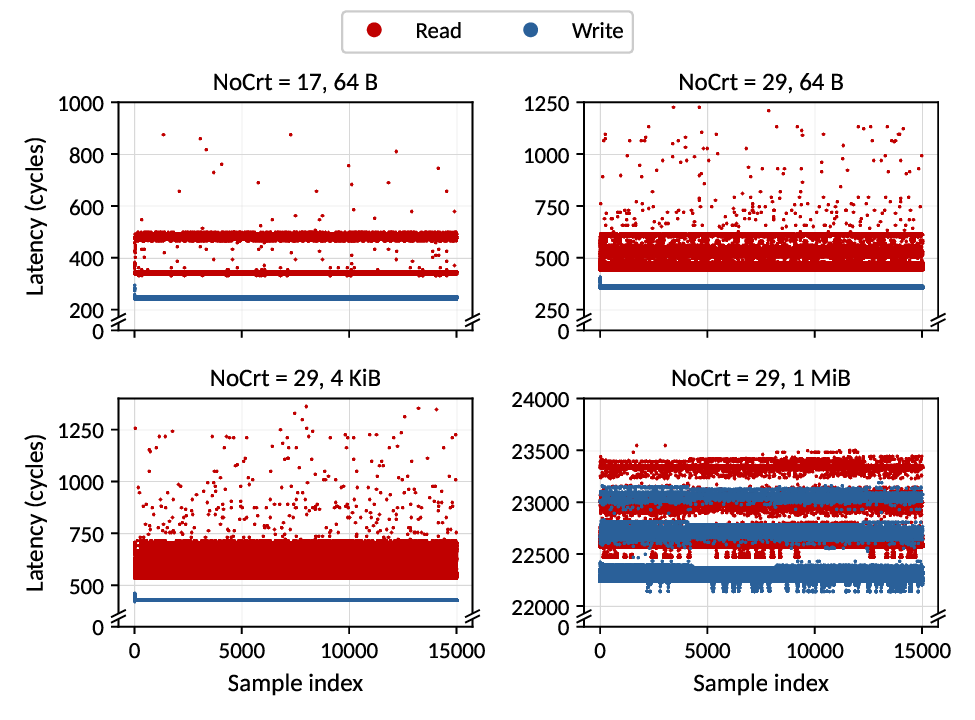}
  \caption{DRAM access latency for four combinations of round-trip distance and
  data size.}
  \label{fig:dram_rw_4096}
\end{figure}

Next, we benchmarked the cost of DRAM access from our reference core. All eight
DRAM partitions were accessed through each of their three DRAM cores on both
NoCs. In this case, the minimum data size was increased to $64$~B to match the
DRAM alignment. Since the DRAM cores are arranged in two columns, the $NoCrt$
between any Tensix core and any DRAM core can only be one of two values: $17$,
when they both have the same $y$ coordinate and differ only in the $x$ coordinate or
$29$, when they differ in both $x$ and $y$ coordinates.
Figure~\ref{fig:dram_rw_4096} shows the read and write latencies for all
accesses with $NoCrt = 17$ and $29$ at $64$~B, and with $NoCrt = 29$ at $4096$~B
and $1$~MiB.

{\bf Observation 2:} DRAM reads show higher variability than remote L1 accesses.

Comparing the results in Figures~\ref{fig:l1_rw}
and~\ref{fig:dram_rw_4096}, it is clear that DRAM reads exhibit a higher degree
of variability than L1 accesses. As L1 accesses also involve the NoC and are
still predictable, the source of the variance must be the DRAM. Interestingly,
this variance does not seem to extend to DRAM writes as their scatter plot
follows a similar pattern to L1 writes. A likely cause for this is the
difference in how the DRAM controller handles reads and writes.

A read response
requires data to be read from the DRAM, while a write acknowledgment can be sent
as soon as the request is placed in the write buffer of the controller. Hence
the read latency includes the timing behavior of the DRAM while a write is
dominated by the traversal time on the NoC and is hence predictable. When the
size of the DRAM write is increased past $16$~KiB, similar variability is
observed in the DRAM write latency plot.

\begin{figure}[htb]
  \centering
  \includegraphics[width=0.95\linewidth]{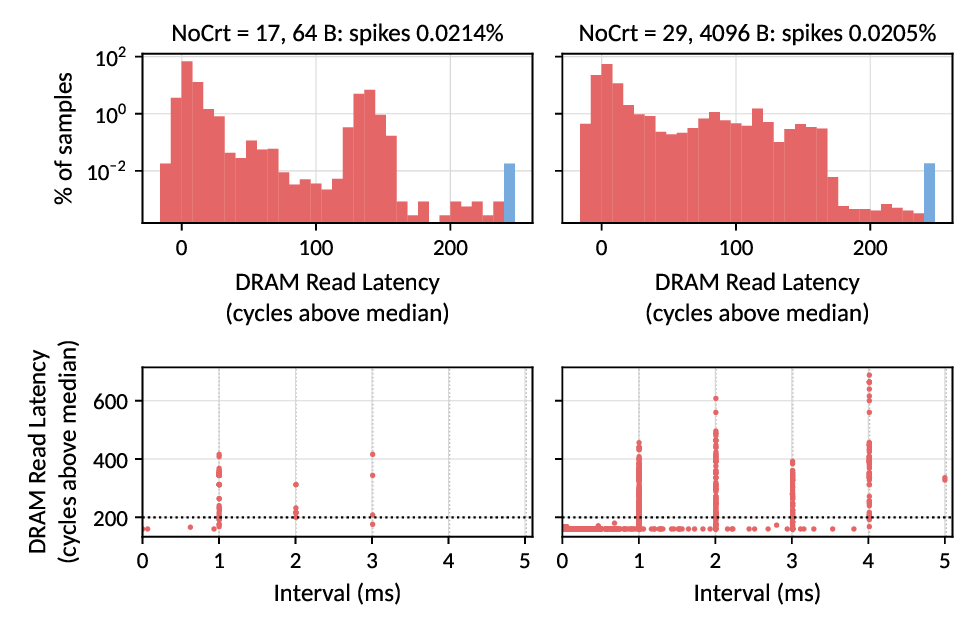}
  \caption{Analysis of DRAM spikes.}
  \label{fig:dram_spikes}
\end{figure}

Since we use observed maxima to
evaluate our model in Section~\ref{sec:exp_eval}, we investigated the DRAM read
latency for the $4096$~B case
of Figure~\ref{fig:dram_rw_4096}. Observe there is a dense band of values below $750$ cycles, representative of the typical
behavior of the DRAM, with a sparse collection of latency spikes that reach up to $1363$ cycles. The top half of
Figure~\ref{fig:dram_spikes} shows a histogram of DRAM read latencies at two
different ($NoCrt$, size) cases: $(17, 64)$ and $(29, 4096)$. Latencies are
grouped based on their difference from the median, which is why there are some
values to the left of 0. The blue bar at the end denotes the catch-all bin of
the histogram and contains all values that are more than $240$~cycles higher
than the median. Note that the $y$-axis uses the log scale, indicating the
relative rarity of values that are several hundred cycles over the median.

We initially classify all
values more than $160$ cycles over the median as spikes, due to their rapid dip in frequency (as seen in the top half of Figure~\ref{fig:dram_spikes}). We then determine the
interval between each pair of consecutive spikes and plot them in a scatter
plot, shown in the bottom half of Figure~\ref{fig:dram_spikes}. An interval has
one spike before it and one after it. We associate each interval with the
minimum of these two latency spikes, and plot the interval on the $x$-axis, and
its associated spike value on the $y$-axis, as the difference between it and the
median. 

{\bf Observation 3:} DRAM accesses are affected by a periodic event that
increases the latency by several hundred cycles. 

As seen in Figure~\ref{fig:dram_spikes}, above a revised threshold of median + $200$
cycles, denoted by a black dotted line, all intervals align cleanly at
multiples of $1$~ms. This indicates that there is an event that happens
approximately every $1$~ms that increases the latency of a DRAM operation by
several hundred cycles over the median. A separate experiment, which read from
several DRAM partitions and a remote L1 at the same time, showed that the latency spikes
occur at the DRAM partition rather than at the DRAM core. All partitions spike
with the same period but at different phases, so the spikes of different
partitions do not coincide. The reads from the remote L1 never spiked.

The cause of the spikes is not known. The Tenstorrent firmware performs a
dynamic voltage and frequency scaling operation every $1$~ms, but that code does
not appear to interact with the DRAM controllers. As the DRAM firmware is not
open, we leave a detailed analysis of the DRAM controller to future work. The
threshold of median + $200$ cycles applies only to sizes up to $4096$~B; other
sizes require the same analysis.

\subsection{Compute}\label{subsec:compute_bm}

The DM baby RISC-V cores issue instructions to the NIU and continue. However,
the Metalium stack provides barriers that allow the DM cores to wait until a NoC
transaction completes, facilitating the aforementioned remote memory access
analysis. However, the same is not possible with the baby RISC-V cores involved
in compute. The compute kernel is compiled into three binaries that run
concurrently on the Unpack, Math and Pack cores, issuing instructions to the
unpacker, FPU or SFPU and packer, respectively. While there are semaphores that
allow a compute baby RISC-V core to wait until all instructions issued to its
coprocessor finish, the documentation states that the wait time might be longer
than necessary. Hence, to simplify our analysis, we model all three compute
baby RISC-V cores as a single unit. 

One further aspect that is relevant for timing analysis is the interaction
between the reader, compute and the writer kernels. The typical pathway is
through circular buffers which we model in the next section. However, they also
contend for L1 ports. On measuring the total time of the compute kernel with and
without contention we found that the time under contention is not strictly
higher. However, to account for it, in our evaluation in
Section~\ref{sec:exp_eval}, we benchmark with and without contention and choose
the maximum of the two values. 

\subsection{NoC Model Parameters}\label{subsec:noc_model_param}

The NoCs on the Blackhole use wormhole routing, and expose counters at each
router that report the number of outgoing packets along each axis. The registers
are readable both from the host and from the Tensix cores nearest to the router.
These counters, read alongside timed data transfers, were used to determine the
timing parameters of the NoC.

First, we checked whether there was any background traffic on the board in the
absence of workloads. Board-wide router counters show no background traffic on
NoC1 over windows of up to $1$~s, but there was some traffic on NoC0. The
latencies obtained in the remote memory access microbenchmarks showed no clear
difference between latencies on NoC0 and NoC1. Regardless, for the evaluation of
our model, we account for this background traffic by considering the maximum
latency across both NoCs. 

The next aspect that affects workloads is dispatch traffic. In fast dispatch
mode, Tensix dispatch cores ($x = 16$) place kernel binaries in the L1 memories
of the host-specified Tensix cores, and issue a go signal to all Tensix cores to
initiate their execution. When a workload is spread across multiple cores, one
of them might start issuing NoC transactions while the go signal is still on the
NoC and being received by other cores, affecting its latency. On reading the
counters, it is found that all launch traffic is restricted to NoC0. Therefore,
in all our experiments, we map readers to NoC1, ensuring that they avoid the
dispatch traffic. Since writers can only start issuing NoC transactions after
the reader and compute kernels produce output, and because the duration of the
dispatch traffic is much smaller than the sum of the read and compute time, we
place writers on NoC0, avoiding the need to account for NoC contention due to
dispatch. 

However, in the case of multi-core workloads, different cores can contend with
each other for the NoC. To bound the NoC delay of a transfer in a multi-core
program, we use the Buffer-Aware Timing Analysis (BATA) of Giroudot and
Mifdaoui~\cite{BATA}, the closest existing analysis to the Blackhole's NoCs. It
models wormhole routing with finite buffers, including indirect blocking: a flow
can be stalled by another flow that never shares a link with it, through an
unrelated packet that shares links with both of them. Using the per-router counters, we
derive the parameters that BATA requires. 

A router passes a flit to the next router with a latency of $9$ cycles, and has
a throughput of one flit per cycle. On a saturated link, a packet holds the link
for its flits plus at most $13$ cycles: a $16$~KiB packet, which has $1$ header
flit and $256$ data flits, takes up to $266$ cycles as a write and $270$ as a
read response. Each router has a per-VC input buffer of size $16$. Since the
buffers are private to VCs, a full buffer on one VC does not affect traffic on
others.

Metalium assigns read requests and writes statically to VC $1$. Read
responses and write acknowledgments use VC $6$ or $7$, with the routers choosing
between the two at each hop. When a router receives multiple packets belonging
to the same VC, it uses round-robin, one packet at a time, while packets on
different VCs share a link flit by flit. While the NoC does support packet
priorities, where a higher-priority packet is chosen ahead of a lower-priority
packet when they both belong to the same VC, we do not use them as they can
cause starvation. 

Using the aforementioned values in BATA still results in a pessimistic bound due
to the differences between the NoC models. BATA assumes arbitrary arbitration
within a VC rather than round-robin. Also, BATA only models fixed priorities
between VCs, whereas the Blackhole's VCs have the same priority and share a link
equally. We place requests and writes in one class and responses and
acknowledgments in another, and pessimistically consider each class to be of
higher-priority when analyzing a flow of the other.

\section{Timing Model}\label{sec:timing_model}

\begin{figure}[htbp]
  \centering
  \includegraphics[width=0.7\linewidth]{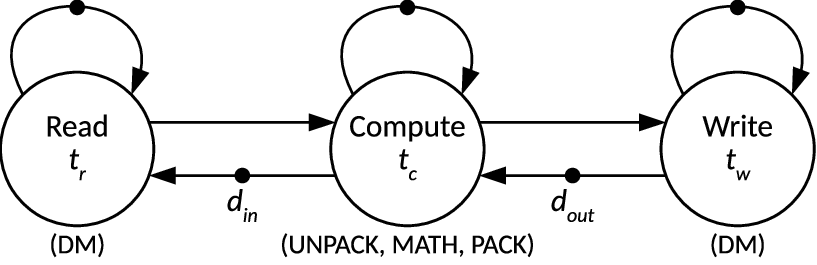}
  \caption{Tensix execution as a Homogeneous Synchronous Dataflow Graph.}
  \label{fig:hsdfg}
\end{figure}

Since execution on a Tensix core includes a read, a compute and a write kernel
which communicate through circular buffers, it can be modeled as a synchronous
dataflow graph~\cite{sdf}. The kernels are actors, and the buffers are edges. To
simplify the analysis, we choose the granularity at which the graph turns into a
Homogeneous Synchronous Dataflow Graph (HSDFG). The execution time of each actor
is chosen such that one firing of the graph produces one output tile. The
resultant HSDFG is shown in Figure~\ref{fig:hsdfg}. Each node is an actor,
labeled with its worst-case firing time ($t_r$, $t_c$ or $t_w$) and the baby
RISC-V cores that run it: one DM core each for the read and
write actors, and the Unpack, Math and Pack cores for the compute actor.
Following prior work~\cite{sdfg_throughput, sdfg_buffer}, each circular buffer
is modeled by a forward edge with {\em no} initial tokens as they start empty,
and a backward edge with $d_{in}$ or $d_{out}$ initial tokens (the sizes of the
input and output buffers in tiles) representing the available slots in the
buffer. If a computation takes $N_r$ inputs, there are $N_r$ input buffers, each
of size $d_{in}$ tiles. The read and compute kernels push and pop one tile in
each input buffer at a time, so they all fill and drain together. Hence we use a
single edge to represent them, with a token standing for one tile of each input.
The backward edge is needed because, if a read kernel produced $d_{in}$ tokens
but the compute did not consume any of them, the buffer between them would be
full, resulting in the read kernel not being able to make progress.

As the execution of the kernels is only blocked by the buffer dependencies
between them, the actors in the HSDFG are self-timed, i.e., they fire as soon as
they have enough data. In a typical self-timed HSDFG, concurrent firings of an
actor are allowed. However, the baby RISC-V cores do not support this. Hence, a
self-loop is added to all actors to disallow concurrent firings. One initial
token is included in this self-loop to allow the first firing.

\subsection{Task Model}\label{subsec:task_model}

Let $\tau$ be a task, defined as all kernels that belong to a single Metalium
program and run on $m \geq 1$ Tensix cores. If a task runs on more than one
Tensix core, it replicates the same read, compute and write kernels across all
its cores. Table~\ref{tab:task_params} lists the parameters that identify it.

\begin{table}[htbp]
  \caption{Parameters of a task $\tau$.}
  \label{tab:task_params}
  \centering
  \small
  \begin{tabular}{|c|p{0.58\linewidth}|}
    \hline
    {\bf Parameter} & {\bf Meaning} \\
    \hline
    $\Pi = \{\pi_0, \dots, \pi_{m-1}\}$ & Tensix cores that $\tau$ runs on \\
    \hline
    $In_i$, $Out_i$ & Sets of endpoints that $\pi_i$ reads from (one per concurrent read)
    and writes to \\
    \hline
    $dtype$ & Datatype of inputs and outputs \\
    \hline
    $d_{in}$, $d_{out}$ & Size of each input buffer and of the output buffer, in
    tiles \\
    \hline
    $N_r$, $k$ & $k$ groups of $N_r$ concurrent reads per firing \\
    \hline
    $N$ & Output tiles per core, hence firings \\
    \hline
    $Op$ & (operation, tiles) pairs performed per output tile \\
    \hline
  \end{tabular}
\end{table}

Each core $\pi_i$ reads its inputs from the endpoints in $In_i$ and writes its
outputs to $Out_i$, each of which is a DRAM endpoint or a remote L1. The
endpoints are specific to each core, while the other parameters are common to
all cores of $\tau$. The cores of $\tau$ do not exchange data with each other.
However, since each core has its own endpoints, its delays can differ from those
of the others, so the worst-case execution time of $\tau$ is the maximum over
all its cores. The datatype $dtype$ determines the size of every transfer and
the cost of computation.

Consider the multiplication of an $X \times Y$ matrix by a $Y \times Z$ matrix,
where all dimensions are in tiles. The output has $XZ$ tiles, and computing each
of them takes $Y$ pairs of input tiles. Since each firing of the graph produces
one output tile, each core fires a total of $N$ times; in this example, $N =
XZ$. This framing determines what each actor does in a firing. 
\begin{itemize}
  \item The writer writes one tile per firing.
  \item $Op$ is the sequence of all operations the compute kernel performs to
  produce one output tile. Each item gives an operation and the number of tiles
  it covers before a single Pack instruction. In this example, $Op$ would be
  ({\tt matmul}, $Y$), which multiplies $Y$ tile pairs, adding the results in
  one register, before packing them as one output tile. 
  \item The reader has to supply all the tiles that these operations consume,
  which can be multiple per input. So, in each firing, it issues $k$ groups of
  $N_r$ reads; $k = Y$ and $N_r = 2$ in this example. The reads of a group are
  in flight concurrently, each from its own endpoint in $In_i$, and the reader
  waits for all of them to finish before issuing the next group. Thus, a firing
  reads $kN_r$ tiles.
\end{itemize}

\subsection{Assumptions}\label{subsec:assum}

The following are assumptions of our model, distinct from any restrictions
imposed by the hardware.

{\bf Host:} Inputs are assumed to be converted to a tile format by the host.
After sending the program to the device it is assumed that the host does not
interact with the device, i.e., does not cause any additional NoC traffic.

{\bf Within a Tensix Core:} The reader and writer of a core always use different
NoCs and access different DRAM partitions or remote L1s ($In_i \cap Out_i =
\emptyset$), so we do not consider any contention between them. The analysis
does not depend on which NoC each one uses. 
Each DM core is mapped to one NoC, and hence to one NIU. In a group of $N_r$
reads issued concurrently by a read kernel, each transfer accesses different
DRAM partitions or L1s. Their requests and responses may still share links, but
we assume that the resulting delay is smaller than the issue and NIU times that
the read term charges for a group, so we do not consider it separately.
Contention for the L1 ports among the read, compute and write kernels is folded
into the compute term. Every tile is smaller than the maximum packet size of
$16$~KiB, so each read or write is a single packet rather than multiple packets
in one transaction. A reader waits for its group of reads, and a writer for its
write, to finish before issuing the next.

{\bf Across Tensix Cores:} Different cores of a task only contend with each
other for the NoC, as each of them has a distinct set of input and output
endpoints. No DRAM partition or remote L1 is accessed by more than one core. The
cores of $\tau$ start on a go signal, which we measured to be brief. 
We therefore assume that the start skew between the cores of $\tau$ is
negligible. We assume that $\tau$ is the only task using the NoC while it runs.
While the compute kernel is unaffected, the reader and writer face interference
on the NoC from the other cores of $\tau$, which we account for using
BATA~\cite{BATA}.

\subsection{Single-Core Worst-Case Execution Time}\label{subsec:wcet}

We first build the firing times of the three actors, and then compose them into
the worst-case execution time of a single core of $\tau$, in the absence of NoC
contention.

{\bf Read Term ($t_r$):} The reader moves data from DRAM or remote L1 into the
input buffers in local L1. 
We use groups of $N_r$ reads because that is the largest number of concurrent
reads that do not contend with each other at the endpoints, as each input is in
its own DRAM partition or remote L1. 

Let $\ell_r(j)$ be the worst-case latency of a read of input $j$. It covers the
issue of the request, the traversal of the NoC, the access at the endpoint and
the delivery of the response. As seen in Section~\ref{subsec:rma_bm}, it is
determined by the size of the transfer, which $dtype$ fixes as each read is one
tile, and by the $NoCrt$ of the endpoint. $\ell_r(j)$ is the maximum over every
endpoint with that $NoCrt$ and over both NoCs.

Since multiple reads are in flight concurrently, they do not each pay $\ell_r$.
As they target different memory locations, the time they spend on the NoC and at
the endpoint overlap. The only parts that serialize are (1) at issue of requests
as they are sent one at a time by the DM core and the NIU, and (2) the response
delivery, as they again queue behind one another at the NIU. Let the issue time
of each request be $t_{issue}$. Let the time that the response of input $j$
spends at the NIU be $t_{resp}(j)$. Removing both from the latency of a single
read leaves the time spent in the NoC and at the endpoint:
\begin{equation}
  D_j = \ell_r(j) - t_{issue,min} - t_{resp,min}(j) \label{eqn:D}
\end{equation}
We subtract their smallest values, $t_{issue,min}$ and $t_{resp,min}(j)$, so
that $D_j$ is not underestimated. Let $D_{max} = \max_j D_j$ and $t_{resp} =
\max_j t_{resp}(j)$.

\begin{figure}[htbp]
  \centering
  \includegraphics[width=0.9\linewidth]{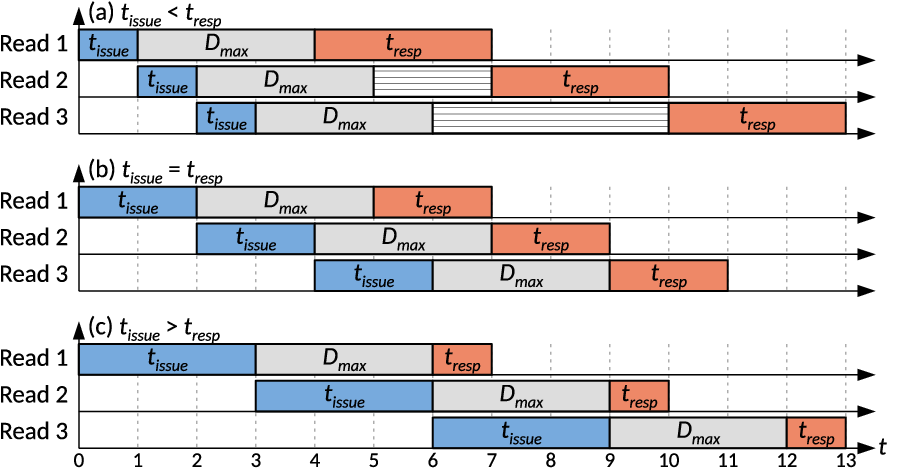}
  \caption{Timing of a group of $N_r = 3$ reads, each taking the full
  $D_{max}$, with time in arbitrary units. Hatched blocks show a response
  waiting at the NIU.}
  \label{fig:read_group}
\end{figure}

The following describes the worst case for a group, shown in
Figure~\ref{fig:read_group}. Each request is issued $t_{issue}$ time after the
previous request, every read takes the full $D_{max}$, and the responses also
spend the full $t_{resp}$ each at the NIU. 
If $t_{issue} < t_{resp}$, the responses queue at the NIU. The first
is delivered at $t_{issue} + D_{max} + t_{resp}$, and each of the
others $t_{resp}$ after the one before it. If $t_{issue} \geq
t_{resp}$, the NIU is free when each response arrives. The group then
ends when the last request, issued at $N_r \cdot t_{issue}$, has spent
$D_{max}$ on the way and $t_{resp}$ at the NIU.
In both
cases, the first read takes the whole path, and the subsequent $N_r - 1$
reads cost only an additional issue or delivery, whichever is higher:
\begin{equation}
  t_{group} = D_{max} + t_{issue} + t_{resp}
               + (N_r - 1)\max(t_{issue},\, t_{resp}) \label{eqn:t_group}
\end{equation}
Here, $t_{issue}$ is the largest time between two consecutive issues, which
includes any instructions the kernel runs between them, and $t_{resp}(j)$ is the
largest time that a response of input $j$ occupies the NIU.
The reader finishes one group before issuing the next, so the firing time of the
read actor is:
\begin{equation}
  t_r = k \cdot t_{group} + e_r \label{eqn:t_r}
\end{equation}
where $e_r$ is the reader's kernel overhead, e.g., cost of any loads, stores or
loop operations performed, per firing. All kernel overheads will be discussed later in this section.

{\bf Compute Term ($t_c$):} Since we model the Unpack, Math and Pack cores as a
single unit, the firing time of the compute actor spans all three.
The firing time has to cover the prefill, steady-state and drain phases of this
three-stage pipeline, so we use its maximum over them. For read-bound workloads,
the pipeline drains and refills at every firing, so this maximum is required.
For compute-bound workloads, the pipeline reaches its steady state, so the
maximum is pessimistic.

Let $\ell_c^{\,o,\,n}$ be the worst-case time of operation $o$ over $n$ tiles of
datatype $dtype$, including contention for the L1 ports. A firing performs every
item in $Op$, so the firing time of the compute actor is:
\begin{equation}
  t_c = \sum_{(o,\,n)\,\in\,Op} \ell_c^{\,o,\,n} \;+\; e_c \label{eqn:t_c}
\end{equation}
where $e_c$ is any kernel overhead per firing that $\ell_c^{\,o,\,n}$ does not
already include.

{\bf Write Term ($t_w$):} The write kernel issues one write request per firing
to move data from the output buffer in local L1 to a DRAM partition or remote
L1. Let $\ell_w$ be the worst-case latency of a write, which is determined by
the same factors as $\ell_r$. The firing time of the write actor is:
\begin{equation}
  t_w = \ell_w + e_w \label{eqn:t_w}
\end{equation}
where $e_w$ is the writer's kernel overhead per firing.

{\bf Kernel Overheads:} The terms above cover NoC transactions and compute
operations. However, each kernel also runs ordinary RISC-V instructions, for
instance to 
read runtime arguments sent by the host.
These are small compared to a transfer or compute, but are not negligible at
small problem sizes, so the model includes a term for them wherever they might
occur: in $t_{issue}$ between two issues, in $e_r$, $e_c$ and $e_w$, per firing,
in $e_{launch}$ once per launch, and in $e_{loop}$ on some firings, such as a
row change in an outer loop of the reader. Some of these are zero for a given
kernel, for example $e_c = 0$ if the buffer traffic of the compute kernel is
already inside $\ell_c^{\,o,\,n}$. These terms depend on how a kernel is
written, and so they should be determined by analyzing its code. We do this by
disassembling the binary of each kernel and counting the instructions on the
path of each term.

{\bf Composing the Terms:} Self-timed execution of the HSDFG in
Figure~\ref{fig:hsdfg} has a transient state followed by a steady state. In the
transient state, the actors are serialized because no forward edge carries an
initial token. Hence the time taken to produce the first output tile is the sum
of the actors' firing times and the reader's per-launch overhead:\footnote{As
all five baby RISC-V cores start together, the compute and write actors
immediately block on buffers that cannot be released until the read actor
completes its first set of transfers. Hence, we only include the read kernel's
per-launch overhead in $e_{launch}$.}
\begin{equation}
  t_{fill} = t_r + t_c + t_w + e_{launch} \label{eqn:t_fill}
\end{equation}

While the first output tile takes the sum of the firing times, each later one is
limited only by the throughput of the slowest part of the graph. This can be a
single actor, or two actors that take turns because they share a small buffer.
In the graph, every such limit is a cycle: a self-loop allows an actor only one
firing at a time, and the backward edge of a buffer allows only as many tiles in
flight as the buffer has slots. A cycle with $d$ tokens and a total worst-case
actor time of $t$ allows at most $d$ firings every $t$ time units, resulting in
a throughput of $d/t$. The interval between firings is the inverse of this
throughput, so the cycle with the largest $t/d$, the maximum cycle
mean~\cite{mcm}, sets the steady-state interval. The graph has five cycles, and each
actor's self-loop has one token corresponding to the firing times, $t_r$, $t_c$ and $t_w$. The cycle
through the input buffer contains the read and compute actors and the $d_{in}$
tokens of its backward edge, which gives $(t_r + t_c) / d_{in}$. Similarly, the
cycle through the output buffer gives $(t_c + t_w) / d_{out}$. Thus, the
steady-state interval between output tiles is the largest of these:
\begin{equation}
  t_{ss} = \max\left(t_r,\; t_c,\; t_w,\;
           \frac{t_r + t_c}{d_{in}},\;
           \frac{t_c + t_w}{d_{out}}\right) \label{eqn:t_ss}
\end{equation}

The transient state ends with the first output tile, and each of the remaining
$N - 1$ follows one steady-state interval later. Thus, the worst-case execution
time of a core in the absence of NoC contention is:
\begin{equation}
  C = t_{fill} + (N-1) \cdot t_{ss} + e_{loop} \label{eqn:C}
\end{equation}

Each term is evaluated per core, using that core's endpoints. All latencies and
costs in the model are maxima. As the model is derived from the architecture and
the programming model, and uses upper bounds for all its terms, it gives a sound
WCET bound for a single-core task in the absence of any NoC or endpoint
contention.

\subsection{DRAM Spikes}\label{subsec:spikes}

Equation~\ref{eqn:t_r}, which is used to derive the WCET of
a single-core task in the absence of contention, uses the worst-case latency of
a read. In the case of a DRAM read, that worst-case latency includes the
periodic event of Observation~3. However, based on experiments, we know that
this only affects a few reads. Hence we tighten the bound in
Equation~\ref{eqn:C} by distinguishing between two maxima: $\ell_r(j)$, the
worst-case read latency including spikes, and $\ell_r'(j)$, the worst-case read
latency excluding them. Note that this distinction only exists for DRAM reads.
For L1 reads, $\ell_r'(j) = \ell_r(j)$, and as Observation~2 shows, the
variability of DRAM reads does not extend to writes at tile sizes. 

We define $t_{group}'$, $t_r'$ and $t_{ss}'$ as in Equations~\ref{eqn:t_group},
\ref{eqn:t_r} and~\ref{eqn:t_ss}, but with $\ell_r'(j)$ in place of $\ell_r(j)$.
This gives the firing time of the read actor, $t_r'$, and the steady-state interval, $t_{ss}'$, 
without spikes. $t_r$ should only be used for firings that are delayed by a
spike. To account for spikes that occur in the first firing, we leave $t_{fill}$
unchanged. If a workload were to experience no other spikes, its WCET would be
given by:
\begin{equation}
  C_0' = t_{fill} + (N-1) \cdot t_{ss}' + e_{loop} \label{eqn:c_0_prime}
\end{equation}

The spikes of Observation~3 recur on each DRAM partition with a period of $P$
cycles, and do not coincide across partitions. The number of DRAM partitions
accessed by a reader is $N_r$. Let the worst-case runtime of a task be $C'$. In
that duration, the task would encounter at most $N_r (\lfloor \frac{C'}{P}
\rfloor + 1)$ spikes. However, one of these is already accounted for in
$t_{fill}$.
This leaves $N_r (\lfloor \frac{C'}{P} \rfloor + 1) - 1$ spikes. So the rest
must occur in the $(N-1)$ steady-state iterations of the task, and the number of
firings they delay is at most:
\begin{equation}
  n_s(C') = \min\left(\left(N - 1\right),\;
        \left(N_r \left\lfloor\frac{C'}{P}\right\rfloor + N_r - 1\right)\right)
        \label{eqn:n_s}
\end{equation}

A firing delayed by a spike uses $t_r$ instead of $t_r'$, which lengthens its
worst-case interval from $t_{ss}'$ to at most $t_{ss}$. Each of these firings
therefore costs at most $(t_{ss} - t_{ss}')$ on top of $C_0'$. As a longer
runtime can encounter more spikes and each spike lengthens the runtime, i.e.,
$n_s$ and $C'$ depend on each other, $C'$ is recursive. We solve it using
fixed-point iteration, which terminates when the next value of $C'$, here referred to as $C'_{next}$, no longer increases from its previous value, $C'_{prev}$, over a maximum number of firings, $n_s=N - 1$. Starting from $C'_{prev} = C_0'$, we repeat the following
until $C'_{next} = C'_{prev}$:
\begin{equation}
  C'_{next} = C_0' + n_s(C'_{prev}) \cdot (t_{ss} - t_{ss}') \label{eqn:c_prime}
\end{equation}

Using $t_r$ for each delayed firing charges a spike to all $k$ read groups,
which is pessimistic when $k > 1$. As $n_s$ is at most $(N - 1)$, $C' \leq C$
and for L1 inputs, $C' = C$. $C'$ is also a sound bound as long as the spikes
behave as in Observation~3. Unlike $C$, it therefore relies on measured behavior
rather than documented architectural behavior.

\subsection{Multi-Core Worst-Case Execution Time}\label{subsec:multi-core}

While the reader and writer of a multi-core task do not contend within a core,
they contend across cores for the NoC as all readers are on NoC1 and all writers
on NoC0. Hence, we extend the single-core task bound by accounting for the
resulting delay using BATA~\cite{BATA}, which gives the delay of one packet of a
flow $f$ as:
\begin{equation}
  D_f^{P_f} = \frac{\sigma_f}{R_f} + T_{hp} + T_{sp} + T_{lp} + T_{IB} + T_{P_f}
  \label{eqn:bata}
\end{equation}
where $\sigma_f$ is the burst of $f$, $R_f$ is the rate left for $f$ on the
busiest link of its path, and $T_{P_f}$ is its base latency. $T_{hp}$ and
$T_{sp}$ are the delays from packets of higher or same priority ($hp$ or $sp$) flows sharing links with $f$, in other VC classes and in its own. $T_{lp}$ is the delay from a lower-priority packet already in
progress at each router of its path, and $T_{IB}$ is the delay from indirect
blocking.

The NoC parameters derived in Section~\ref{subsec:noc_model_param} and the flows
of a multi-core workload are used in Equation~\ref{eqn:bata}. Each read is made
of two flows: a request and a response, and each write consists of two flows:
the write and its acknowledgment. The route of a flow is determined by the
core, endpoint and the NoC. A kernel can have the $N_r$ reads of a group in
flight together, but each read has its own endpoint and hence its own flows, and
the kernel waits for the whole group, or write, to finish before issuing the
next. Each flow therefore has a burst of at most one packet in flight, and its rate is one packet per the shortest time between two of its
transfers.

Equation~\ref{eqn:bata} gives the whole delay of a packet, but $\ell_r$ and
$\ell_w$ already include its traversal of the NoC ($T_{P_f}$) and its
transmission without contention ($L_f / R$, for a packet of $L_f$ flits on a
link of rate $R$). The delay that contention adds to a flow is therefore:
\begin{equation}
  \Delta_f = D_f^{P_f} - T_{P_f} - \frac{L_f}{R} \label{eqn:delta_f}
\end{equation}
This gives $\Delta_{req,j}$ and $\Delta_{resp,j}$ for the request and response
of read $j$ of a group, and $\Delta_{wr}$ and $\Delta_{ack}$ for a write and its
acknowledgment. A group is delayed by its slowest read. The requests of a group
leave a core one after another on the same VC, so if another core's traffic
stalls a request, the requests behind it wait too. The delay of a read therefore
includes the delays of all the requests issued before it, as well as its own.
The delays of a group and of a write are thus:
\begin{align}
  \Delta_{group} &= \max_j \left(\sum_{i \leq j} \Delta_{req,i} + \Delta_{resp,j}\right) \label{eqn:d_group}\\
  \Delta_{write} &= \Delta_{wr} + \Delta_{ack} \label{eqn:d_write}
\end{align}
We add $\Delta_{group}$ to $t_{group}$ and $t_{group}'$, including any group
read inside $e_{launch}$, and $\Delta_{write}$ to $t_w$, and then evaluate the
equations above with the endpoints of each core $\pi_i$ to get $C_i$ and $C_i'$.
The WCET of a multi-core task $\tau$ is then obtained as
the highest single-core WCET among its $m$ cores: $C = \max_{i=0}^{m-1} C_i$,
and its tightened form is $C' = \max_{i=0}^{m-1} C_i'$.

\section{Experimental Evaluation}\label{sec:exp_eval}
In this section, we evaluate the model of Section~\ref{sec:timing_model} across
configurations, workloads and core counts. All measurements were taken on a
Tenstorrent Blackhole p150a, hosted by an
Intel Core i7-13700K processor, running Ubuntu 22.04.5 (Linux
6.8.0), with Metalium v0.75.0 and firmware version 19.8.1. 
Times are reported in cycles, whose frequency is pinned at $1350$~MHz using host
hints and never drops below $1343$~MHz. 

We use observed maxima in the equations presented in
Section~\ref{sec:timing_model}. Each DRAM and L1 latency was obtained by taking
the maximum across $15{,}000$ samples. All other microbenchmarks and workloads
run for $10{,}000$ iterations. As these are observed maxima rather than proven
upper bounds, the values of $C$ and $C'$ reported here are WCET
estimates~\cite{wcet_problem}. We compare them with the observed worst case, the
largest total time over all iterations.

\subsection{Generality Across Configurations}
\label{subsec:diff_config}

We first select a single-core workload and vary its configuration. The workload
is a 
matrix multiply of $X \times Y$ tiles by $Y \times Z$ tiles, so $N = XZ$, $k =
Y$ and $N_r = 2$. We vary the datatype ({\tt Bfp8\_b}, {\tt Float16\_b} or {\tt
Float32}), the input and output circular buffer depths ($1$, $2$, $4$ or $8$
each), and the problem size. Eight problem sizes $X{\times}Y{\times}Z$ were
chosen to cover square shapes ($1{\times}1{\times}1$ to
$20{\times}20{\times}20$), a non-square shape ($4{\times}16{\times}32$), a
write-heavy shape with no accumulation ($16{\times}1{\times}16$), and a
deliberately read-heavy shape ($1{\times}32{\times}1$), which issues $32$ read
groups for only one firing of its HSDFG. This gives a total of $3 \times 4
\times 4 \times 8 = 384$ configurations. 

We then run each of the above configurations for different locations
of three operands (two input and one output) based on memory type and $NoCrt$
class. In the case of the DRAM, there are only two possible $NoCrt$ classes: 17
and 29. We consider five combinations of round-trip hop distances for the three
operands: (17,17,29), (17,29,17), (17,29,29), (29,29,17) and (29,29,29). This
results in a total of $5 \times 384 = 1920$ DRAM configurations.

On the L1, all three $NoCrt$ classes are possible. Hence with three
operands and three possible $NoCrt$ classes, we get a total of $3^3 = 27$
combinations, resulting in $27 \times 384 = 10{,}368$ L1
configurations.\footnote{Two of the eight problem sizes were reduced to fit
within the $1.5$~MiB of Tensix core's L1: $20{\times}20{\times}20 \rightarrow
18{\times}18{\times}18$ and $4{\times}16{\times}32 \rightarrow
4{\times}15{\times}23$.}

\begin{figure}[htbp]
  \centering
  \includegraphics[width=0.8\linewidth]{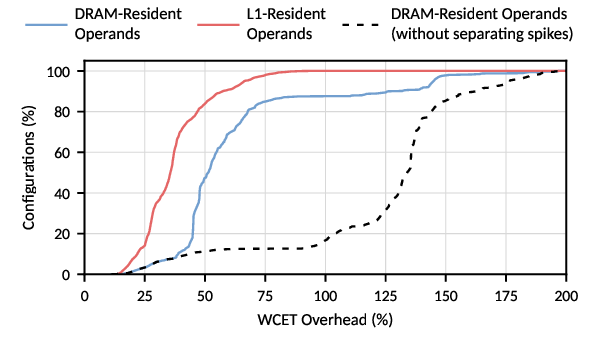}
  \vspace{-0.15in}
  \caption{WCET overhead across different configurations of matrix multiply.}
  \label{fig:wcet_cdf}
\end{figure}

Figure~\ref{fig:wcet_cdf} shows the distribution of the WCET overhead. The
estimate is above the observed worst case in every configuration. The difference
between the two DRAM curves is the pessimism that Equation~\ref{eqn:c_prime}
avoids. Without it, $C$ charges a spike to every read, which puts most DRAM
configurations above $100\%$, although only a few reads actually encounter one.
With it, $C'$ stays below $100\%$ in all configurations except those of
$1{\times}32{\times}1$. That shape is problematic by design: its graph only has
a single firing, so there is no steady state to tighten, and $t_{fill}$ charges
a spike to each of its $32$ read groups.

Matrix multiplication is a read-dominated workload. Hence, the read term decides
the steady-state interval throughout the DRAM configurations,
directly or through the input buffer's cycle when $d_{in} = 1$. Even after
tightening by accounting for DRAM spikes, it applies the maximum
latency to every read. L1 configurations result in much lower WCET
overhead as their maximum latency is much less pessimistic. An L1 read has no
spikes and its maximum stays close to its minimum, so the model needs no
tightening ($C' = C$) and the estimate is tighter and below $100\%$ in every
configuration, including $1{\times}32{\times}1$. As the equations are
the same in both L1 and DRAM cases, much of the pessimism comes from
the unpredictability of the DRAM rather than from the composition.

\subsection{Generality Across Workloads}
\label{subsec:diff_workloads}

\begin{figure}[htbp]
  \centering
  \includegraphics[width=\linewidth]{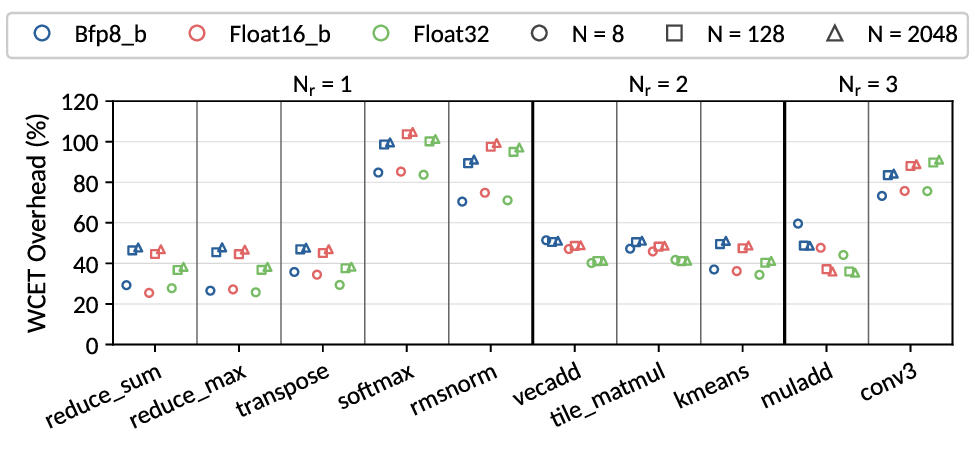}
  \vspace{-0.25in}
  \caption{WCET overhead across different workloads.}
  \label{fig:wcet_workloads}
\end{figure}

Next, we developed ten workloads, drawing from
kernels used in the GPGPU benchmarks of the Vortex
work~\cite{vortex,vortex_github}, together with three primitives from
transformer and neural network inference. Between them, they cover one, two, and
three input streams, and compute stages of one operation or a chain of several.
Each runs at the three aforementioned floating-point datatypes and $N \in \{8,
128, 2048\}$, with both buffers set to size $2$, resulting in a total of $10
\times 3 \times 3 = 90$ configurations. All operands are placed in DRAM
partitions.

Figure~\ref{fig:wcet_workloads} shows that the estimate, factoring DRAM spikes, is again above the
observed worst case in every configuration.
\texttt{softmax}, \texttt{rmsnorm} and \texttt{conv3} are compute-bound and the
most overestimated. Their compute stages chain several operations, and $t_c$
sums $\ell_c^{\,o,\,n}$ for each of them. As $\ell_c^{\,o,\,n}$ is the maximum
across the prefill, steady-state and drain phases of the unpack-math-pack pipeline,
the overall pessimism is high. The remaining seven workloads are read-bound and
result in tighter WCET estimations. Among them, overestimation is lowest at $N_r
= 1$ and slightly higher with more inputs, since a group is charged the worst
case of its slowest read, and the number of spikes that are charged grows with $N_r$.

\subsection{Multi-Core Workloads}
\label{subsec:multi-core_eval}

Next, we apply the model to multi-core workloads to evaluate how it accounts
for NoC delay. 
In all cases, we compare the estimate from the model with the largest observed
WCET across all cores of a workload. 
The values we use in Equation~\ref{eqn:delta_f} are as follows. The coordinates
of the cores and operands determine the routes of different flows. The per-flit
latency and buffer depth are used in the equations of BATA~\cite{BATA}
that build up the terms of Equation~\ref{eqn:delta_f}.
By experimentation, we determined that a packet holds a saturated link for its
flits plus up to $13$ cycles.\footnote{This delay was measured for the maximum
packet size, and hence is likely pessimistic for other sizes. Nevertheless, we
use it for all packets.} Hence, wherever the equations of BATA use the time a
packet holds a link, we use $\left(L_f/R\right) + 13$ instead of
$\left(L_f/R\right)$, including in a flow's burst and rate. 
In Equation~\ref{eqn:delta_f}, we still subtract only $\left(L_f/R\right)$, as
$\ell_r$ and $\ell_w$ do not include the extra $13$ cycles.
For a flow's rate, the shortest time between two of its transfers is at least
the transfer's minimum latency, since a transfer's issue, NoC traversal,
endpoint and NIU times all separate it from the next. We use the minimum
observed value from the latency microbenchmarks at the flow's $NoCrt$ and
transfer size.
This leaves out the time a kernel spends between transfers, so it overstates the
rate of most workloads.

\subsubsection{DRAM-Resident Operands}

\begin{figure}[htbp]
  \centering
  \includegraphics[width=\linewidth]{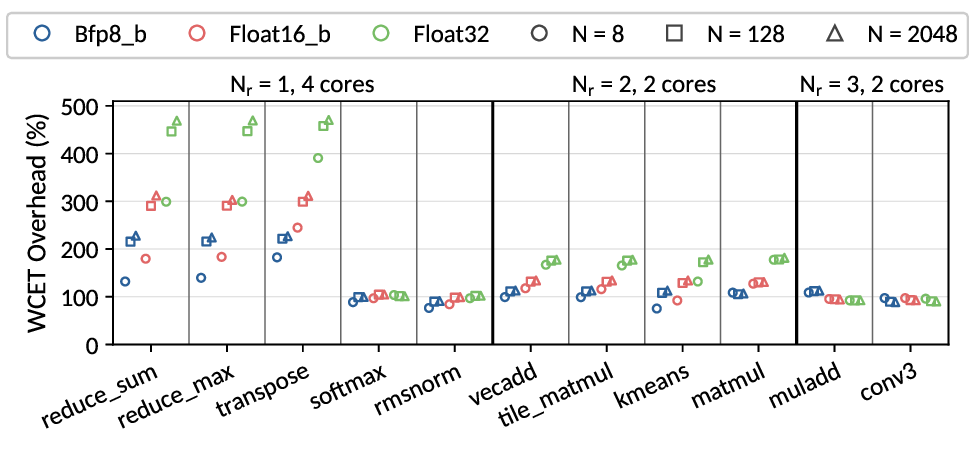}
  \vspace{-0.25in}
  \caption{WCET overhead of multi-core workloads with operands in DRAM.}
  \label{fig:wcet_dram_bata}
\end{figure}

We choose $\lfloor 8 / (N_r + 1) \rfloor$ cores for each of the eleven workloads
discussed in the previous experiments as our task model requires inputs and
output to be in distinct DRAM partitions. We run each workload at three
datatypes and three problem sizes 
resulting in $11 \times 3 \times 3 = 99$ configurations.
Figure~\ref{fig:wcet_dram_bata} shows that the estimate, once again considering DRAM spikes, stays above the observed
worst case in every configuration. Read-bound workloads at $N_r = 1$ show the
most overestimation, which increases with the size of the datatype. As they have
the most cores (4), their read responses share most of their long
routes, and BATA charges a whole packet at every shared link, resulting in a
large $\Delta_{group}$, and as a result a large $C'$.
Compute-bound workloads are close to their single-core trend, as $t_c$ still
dominates the steady state and the NoC term mostly affects the first firing.
Workloads at $N_r = 2$ and $3$ run on fewer cores, resulting in tighter WCET estimates.
The observed worst case itself rises only slightly over the single-core runs, so
most of the overestimation comes from the analysis rather than from contention.

\subsubsection{L1-Resident Operands}

\begin{figure}[htbp]
  \centering
  \includegraphics[width=\linewidth]{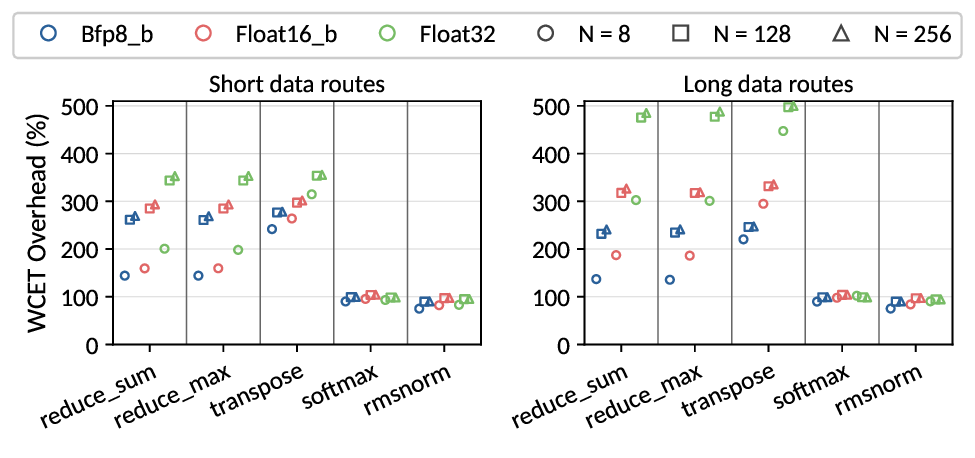}
  \vspace{-0.25in}
  \caption{WCET overhead of 32-core workloads with operands in L1.}
  \label{fig:wcet_l1_bata}
\end{figure}

To increase the core count, we run five $N_r = 1$ workloads on $32$ cores,
each with two dedicated remote L1s for input and output. Problem sizes are
chosen as $N \in \{8, 128, 256\}$ to ensure each operand fits in one L1. To show
the impact of shared links, we chose two configurations of cores and operand
locations: (1) short data routes, where data packets cross only a few links,
minimizing link sharing; and (2) long data routes, where the data
packets go the long way round instead, and data flows of different cores share
links. Figure~\ref{fig:wcet_l1_bata} shows that the estimate is again above the
observed worst case in every configuration.

The compute-bound workloads show negligible difference in WCET overheads for different data routes. For the read-bound workloads, short data routes show
lower overestimation than long ones in general. This is because long routes
share many links and have a higher relative cost. With both placements,
overestimation grows with the size of the datatype, which directly increases the
transfer size and hence the blocking. The observed worst case follows the same
trend, being higher with long data routes and increasingly so with the datatype,
but by far less than the estimate.

\section{Related Work}\label{sec:rel_work}
WCET analysis of accelerators has largely focused on GPUs. Betts and
Donaldson~\cite{gpu_wcet_hybrid} combine static program structure with measured
execution times. Berezovskyi et al. characterize CUDA
kernels by measurement and extreme value theory~\cite{wcet_cuda_kernels} and
compute makespans for threads on one streaming
multiprocessor~\cite{warp_makespan}. At the thread block granularity, Huangfu
and Zhang~\cite{wcet_gpu_warp} derive a static bound under an assumed
warp-scheduling policy, and Jeanmougin et al.~\cite{wcet_thread_block_bound}
bound WCET using a multi-phase representation of warp execution. Bendellaa and
Cucu-Grosjean propose a WCET estimation model for real-time
GPUs~\cite{gpu_wcet_model_wip}. In these works, several aspects of the
hardware's behavior are assumed or only measured, whereas on the Blackhole, we
derive execution patterns and data movement from the architecture and program
flow.

On FPGA SoCs, existing works target the AMD Deep Learning Processor Unit (DPU).
Restuccia and Biondi~\cite{DNN_on_FPGA} derive a response-time analysis from a
profiled execution model and tighten it with a custom FPGA module. Aromolo et
al.~\cite{DNN_on_Multicore_FPGA} give guarantees on multi-core
DPUs using bus budgeting units and limited-preemptive fixed-priority scheduling.
Zhang et al.~\cite{WCET_CNN_on_FPGA} perform a static timing analysis of CNN
inference on multiple DPUs, modeling arbitration on the shared bus and memory
contention. In these systems, the accelerators contend on a shared memory bus.
Cores on the Blackhole instead contend on a wormhole NoC, and we bound that
delay using BATA~\cite{BATA}. On the Versal AI Engines, the nearest commercial
spatial array to the Blackhole, Sprave et al.~\cite{versal_mc} schedule
mixed-criticality tasks using measured WCETs. 

Architecturally, the nearest relative of the Blackhole is the Kalray MPPA, a
clustered many-core with local memory and a NoC. Its NoC provides flow
regulation, and as a result may operate with guaranteed
services~\cite{Kalray_NoC}. Becker et al.~\cite{NoC_partition} partition the NoC
to reduce contention. Hard real-time applications on the MPPA can be temporally
isolated~\cite{Kalray_isol}, and existing response-time analysis covers
synchronous dataflow programs~\cite{Kalray_rta_sdf} and dataflow applications
across clusters, with NoC transfers modeled as tasks~\cite{Kalray_rta_df}. While
the Blackhole's NoC does not provide comparable guarantees, it is designed for
ML workloads, has a fully open software stack~\cite{tt-metal}, and has
relatively open RISC-V hardware~\cite{tt_isa_docs}, making it a more promising
platform for predictable AI acceleration. To the best of our knowledge, no prior
work derives the WCET for this architecture; the vendor's NoC
performance estimator~\cite{tt-npe} gives average-case estimates.

\section{Conclusion \& Future Work}\label{sec:conclusion}
This work describes a WCET analysis of the Tenstorrent Blackhole, a RISC-V-based AI
accelerator with explicitly managed data movement and open software stack.
A task is modeled as a Homogeneous Synchronous
Dataflow Graph, and a WCET is derived for a single-core task given upper bounds on
the model parameters. This time prediction is extended to multi-core workloads by accounting for
inter-core NoC contention. Evaluations show the WCET estimate bounds the observed worst case in every configuration.

In future work, we will extend our timing analysis to include co-running
workloads and a more rigorous study of DRAM latencies. The openness of
the Tenstorrent platform makes it suitable for timing analysis, and warrants
further study by the real-time community.

\section*{Acknowledgment}

Claude Opus 5.5 assisted in code development which will be made publicly
available. The authors verified all generated content and take full
responsibility for its use in this work.

\bibliographystyle{IEEEtran}
\bibliography{references}

\end{document}